\documentstyle[multicol,aps,prb,amssymb,epsf]{revtex}
\begin{document}
\title{Coexistence of ferromagnetism and superconductivity in the hybrid
ruthenate-cuprate compound RuSr$_{2}$GdCu$_{2}$O$_{8}$ studied by muon spin
rotation ($\mu $SR) and DC-magnetization}
\author{C. Bernhard$^{1)}$, J.L. Tallon$^{2)}$, Ch. Niedermayer$^{3)}$, Th. Blasius$%
^{3)}$, A. Golnik$^{1)\ast }$, E. Br\"{u}cher$^{1)}$, R.K. Kremer$^{1)}$,
D.R. Noakes$^{4)}$, C.E. Stronach$^{4)}$ and E.J. Ansaldo$^{5)\#}$}
\address{1) Max-Planck-Institut f\"{u}r Festk\"{o}rperforschung Heisenbergstrasse 1,\\
D-70569 Stuttgart, Germany\\
2) Industrial Research Ltd., P.O. Box 31310, Lower Hutt,\\
New Zealand \\
3) Universit\"{a}t Konstanz, Fakult\"{a}t f\={u}r Physik, D-78434\\
Konstanz, Germany \\
4) Department of Physics, Virginia State University, Petersburg, Virginia\\
23806, USA\\
5)University of Saskatchewan, Saskatoon S7N OWO, Canada}
\date{26.1.99}
\maketitle

\begin{abstract}
We have investigated the magnetic and the superconducting properties of the
hybrid ruthenate-cuprate compound RuSr$_{2}$GdCu$_{2}$O$_{8}$ by means of
zero-field muon spin rotation- (ZF-$\mu $SR) and DC magnetization
measurements. The DC-magnetisation data establish that this material
exhibits ferromagnetic order of the Ru-moments ($\mu $(Ru) $\approx $ 1 $\mu
_{B}$) below {\it T}$_{Curie}$ = 133 K and becomes superconducting at a much
lower temperature {\it T}$_{c}$ = 16 K. The ZF-$\mu $SR experiments indicate
that the ferromagnetic phase is homogeneous on a microscopic scale and
accounts for most of the sample volume. They also suggest that the magnetic
order is not significantly modified at the onset of superconductivity.

\end{abstract}

\allowbreak

\begin{multicols}{2}
%\newpage

\section{Introduction}
\noindent Since the discovery of superconductivity in the cuprate system La$%
_{2-x}$Ba$_{x}$CuO$_{4}$ in 1986 \cite{1} an ever growing variety of high-T$%
_{c}$ superconducting cuprate compounds has been synthesized all of which
contain CuO$_{2}$ planes (some also contain CuO chains) as their essential
structural elements which host the superconducting charge carriers \cite{2}.
Between the CuO$_{2}$ planes are various kinds of layers, typically
NaCl-type, which are insulating and act merely as a charge reservoir. To
date, the ruthenate compound Sr$_{2}$RuO$_{4}$ is the only known layered
perovskite-like system which becomes superconducting even though it does not
contain any CuO$_{2}$ planes or CuO chains \cite{3}. Despite its rather low
transition temperature {\it T}$_{c}$ = 1.5 K the study of its electronic and
magnetic properties has become a very rich and active field of research \cite
{4}. In parallel, the electronic and magnetic properties of the related
ruthenate compounds, such as for example the SrRuO$_{3}$ system which is an
itinerant 4d-band ferromagnet with {\it T}$_{Curie}$ $\approx $ 165 K, have
attracted a great deal of interest \cite{5}.

Another potentially promising and exciting direction of research has been
prompted by the circumstance that the RuO$_{2}$ layers share the same
square-planar coordination and a rather similar bond length with their CuO$%
_{2}$ counterparts. A whole new family of hybrid ruthenate-cuprate compounds
may therefore be constructed whose members consist of different sequences of
alternating RuO$_{2}$- and CuO$_{2}$ layers. Recently, one such a hybrid
ruthenate-cuprate compound, the 1212-type RuSr$_{2}$GdCu$_{2}$O$_{8}$ system
comprising CuO$_{2}$ bilayers and RuO$_{2}$ monolayers, has been synthesized
as single-phase material \cite{6}. A subsequent study of its electronic and
magnetic properties has revealed that this material exhibits electronic
ferromagnetic order at a rather high Curie temperature {\it T}$_{Curie}$%
=133-136 K and becomes superconducting at a significantly lower critical
temperature {\it T}$_{c}$=15-40 K (depending on the condition of preparation
and annealing) [6-8].\ The most surprising observation, however, is that the
ferromagnetic order does not vanish when superconductivity sets in at {\it T}%
$_{c}$. Instead, it appears that the ferromagnetic state remains largely
unchanged and coexists with superconductivity. This finding implies that the
interaction between the superconducting- and the ferromagnetic order
parameters is very weak and it raises the question of whether both order
parameters coexist on a truly microscopic scale. Since the early
investigations of Ginzburg in 1957 \cite{9} the prevailing view is that the
coexistence of a superconducting- (with singlet Cooper pairs) and a
ferromagnetic order parameter is not possible on a microscopic scale since
the electromagnetic interaction and the exchange coupling cause strong
pairbreaking. Indeed, merely based on magnetisation and transport
measurements one cannot exclude the possibility that the RuSr$_{2}$GdCu$_{2}$%
O$_{8}$ samples may be spatially inhomogeneous with some domains exhibiting
ferromagnetic order and others superconducting order \cite{6}. We note that
unambiguous evidence for the occurence of bulk superconductivity in RuSr$%
_{2} $GdCu$_{2}$O$_{8}$ has recently been obtained from specific heat
measurements which reveal a sizeable jump at T$_{c}$ of $\Delta \gamma
\equiv C_{p}/T\approx 0.35$ mJ/g at.K$^{2}$ characteristic of a strongly
underdoped cuprate superconductor \cite{10}. In the following we report on
muon-spin rotation ($\mu $SR) measurements which establish that the
ferromagnetic order is uniform and homogeneous even on a microscopic scale.

The $\mu $SR technique is ideally suited for such a purpose since it
provides an extremely sensitive local magnetic probe and, furthermore,
allows one to reliably obtain the volume fraction of the
magnetically-ordered phase \cite{11}. Here we present the result of a
zero-field muon-spin rotation (ZF-$\mu $SR) study of a RuSr$_{2}$GdCu$_{2}$O$%
_{8}$ sample with {\it T}$_{c}$=16 K and {\it T}$_{Curie}$=133 K which
provides evidence that the magnetic order parameter is spatially homogeneous
and accounts for most of the sample volume. Furthermore, the ZF-$\mu $SR
data establish that the ferromagnetic order is hardly affected by the onset
of superconductivity and persists to the lowest available temperature of the
experiment {\it T}=2.2 K. The ZF-$\mu $SR data can be complemented by
DC-magnetisation measurements which establish the presence of ferromagnetic
order from the observation of a spontaneous magnetization at {\it T}$%
_{Curie} $ = 133 K and of hysteretic isothermal magnetic behavior with a
remanent magnetization. It is shown that the ferromagnetic ordering involves
the Ru magnetic moments with $\mu ($Ru) $\approx $ 1.05(5)\ $\mu _{B}$,
while the larger Gd-moments with $\mu $(Gd$^{3+}$) $\approx $ 7.4(1) $\mu
_{B}$ remain paramagnetic down to very low temperatures. In addition, the
magnetisation measurements indicate an almost complete diamagnetic shielding
effect below T$_{c}$.

\section{Experiment}
\subsection{sample preparation and characterization}
Polycrystalline samples of the 1212-type system RuSr$_{2}$GdCu$_{2}$O$_{8}$
have been synthesized as previously described \cite{8} by solid state
reaction of RuO$_{2}$, SrCO$_{3}$, Gd$_{2}$O$_{3}$ and CuO powders. The
mixture was first decomposed at 960 $%
%TCIMACRO{\UNICODE[m]{0xb0}}%
%BeginExpansion
{{}^\circ}%
%EndExpansion
$C in air. It was then ground, milled and die-pressed into pellets. The
first sintering step took place in flowing nitrogen atmosphere at 1010 $%
%TCIMACRO{\UNICODE[m]{0xb0}}%
%BeginExpansion
{{}^\circ}%
%EndExpansion
$C. This step results in the formation of a mixture of the precursor
material Sr$_{2}$GdRuO$_{6}$ and Cu$_{2}$O and is directed towards
minimizing the formation of SrRuO$_{3}$ \cite{6}. The material was then
reground before it was reacted in flowing oxygen for 10 hours at 1050 $%
%TCIMACRO{\UNICODE[m]{0xb0}}%
%BeginExpansion
{{}^\circ}%
%EndExpansion
$C. This sintering step was repeated twice with intermediate grinding and
milling. Each reaction step was carried out on a MgO single crystal
substrate to prevent reaction with the alumina crucible. Finally the samples
were cooled slowly to room temperature in flowing oxygen. Following this
procedure we have also made a Zn-substituted RuSr$_{2}$GdCu$_{1.94}$Zn$%
_{0.06}$O$_{8}$ sample and a Y $\leftrightarrow $\ Gd cosubstituted sample
RuSr$_{2}$Gd$_{0.9}$Y$_{0.1}$Cu$_{2}$O$_{8}$. X-ray diffraction (XRD)
measurements indicate that all samples are single phase 1212-type material
and give no indication for traces of the ferromagnetic phase SrRuO$_{3}$.
Figure 1(b) displays a representative XRD-spectrum of RuSr$_{2}$GdCu$_{2}$O$%
_{8}$, the plus signs show the raw data and the solid line shows the result
of the Rietveld refinement. The related structure of RuSr$_{2}$GdCu$_{2}$O$%
_{8}$ is shown in Fig. 1(a).

\begin{figure}
\epsfxsize=5.6 cm
\epsfbox{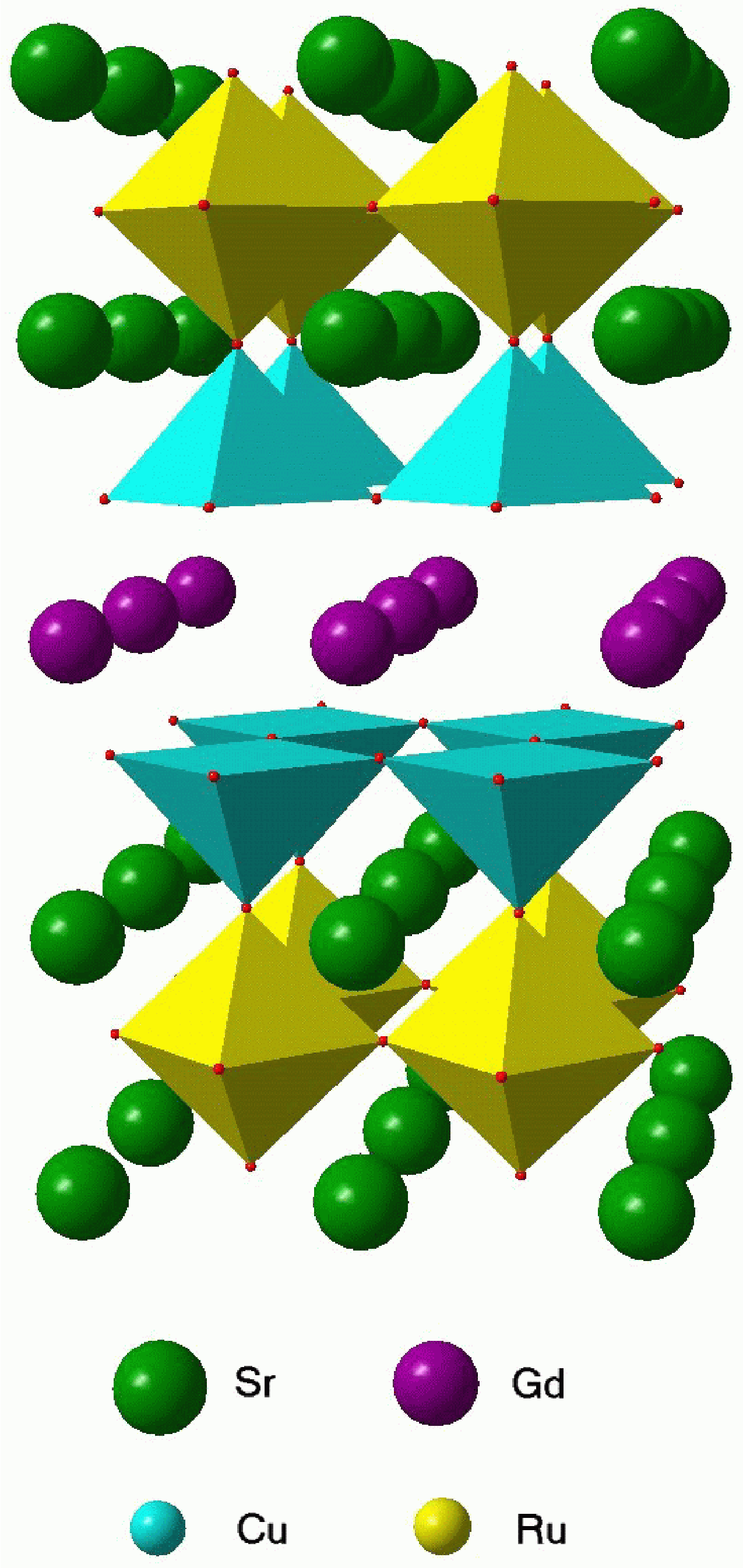}
\epsfxsize=8.4 cm
\epsfbox{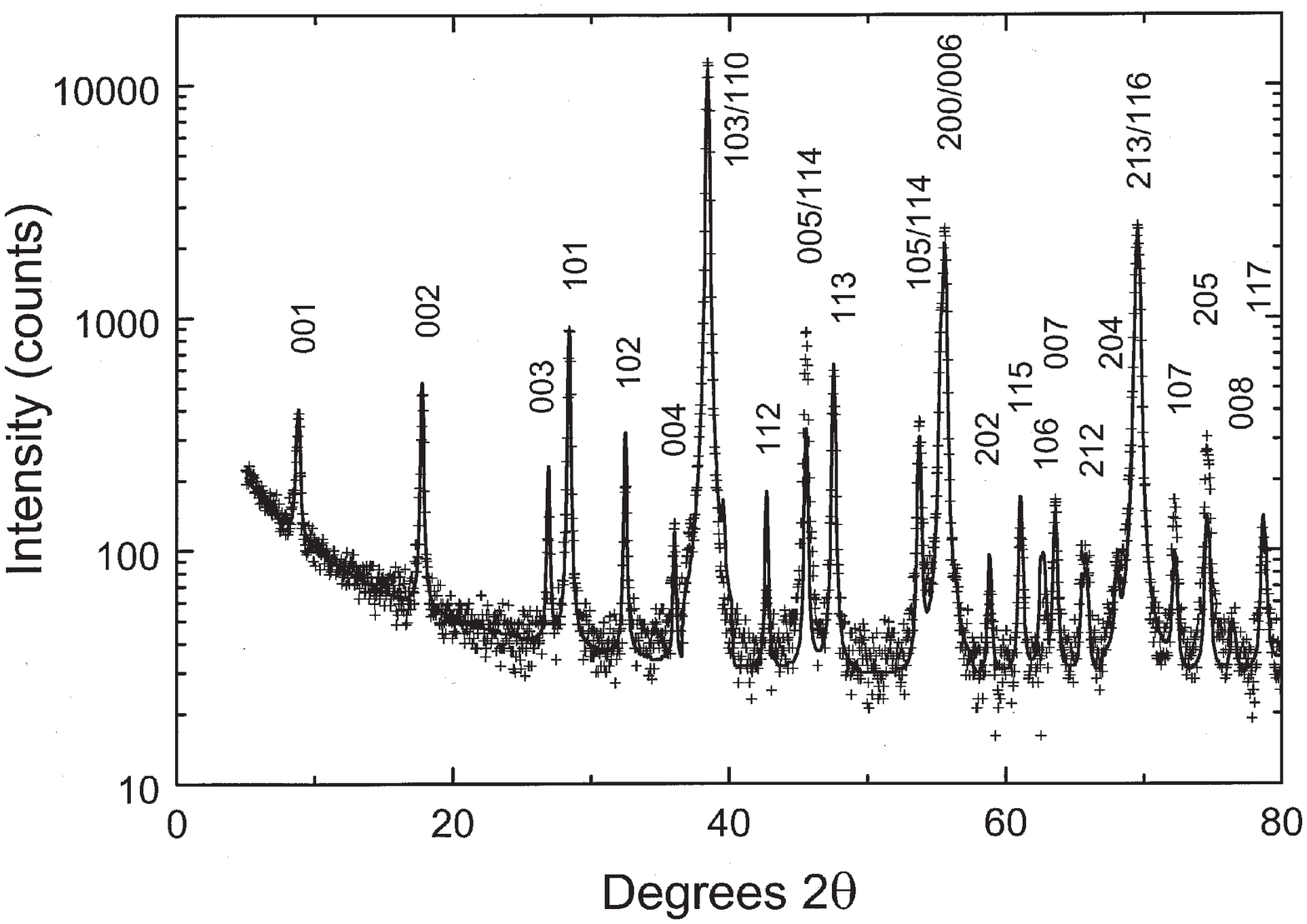}
\caption{(a) The structure of RuSr$_{2}$GdCu$_{2}$O$_{8}$ with the Cu atoms
sited at the centre of the base of the square pyramids and the Ru atoms at
the centre of the octahedra. (b) The X-ray diffraction (XRD) spectrum for a RuSr$_{2}$GdCu$_{2}$O$_{8}
$ sample (Co K-$\alpha $ source). The plus signs (+) are the raw x-ray data
and the solid line is the calculated Rietveld refinement profile for
tetragonal (space group P4/mmm) RuSr$_{2}$GdCu$_{2}$O$_{8}$.}
\end{figure}

The electronic properties of RuSr$_{2}$GdCu$_{2}$O$_{8}$ have been
characterized by measurements of the temperature-dependent resistivity and
thermo-electric power. Representative results are shown in Fig. 2(a) and
2(b) respectively (see also Ref. 8). 

\begin{figure}
\epsfxsize=5.3 cm
\epsfbox{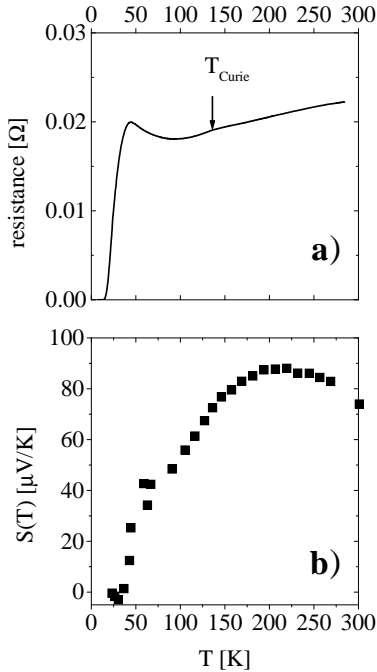}
\caption{(a) Temperature dependence of the resistivity, $\rho $, of RuSr$_{2}$%
GdCu$_{2}$O$_{8-\delta }$. (b) The temperature-dependent thermoelectric
power S(T).}
\end{figure}

The temperature dependence of the
thermo-electric power {\it S}({\it T}) and, in particular, its normal state
value of {\it S}(300 K) $\approx $\ \ 75 $\mu $V/K is rather typical for a
strongly underdoped cuprate superconductor with T$_{c}$ 
%TCIMACRO{\TEXTsymbol{<}}%
%BeginExpansion
\mbox{$<$}%
%EndExpansion
%TCIMACRO{\TEXTsymbol{<} }%
%BeginExpansion
\mbox{$<$}%
%EndExpansion
T$_{c,max}$, consistent with a hole content of {\it p}$\approx $0.07 holes
per CuO$_{2}$ planes and a value of {\it T}$_{c,max}$ of the order of 100 K
[8,12]. The resistivity measurements indicate that the RuSr$_{2}$GdCu$_{2}$O$%
_{8}$ sample exhibits zero resistivity at a critical temperature of {\it T}$%
_{c}$ = 16 K. The precise value of {\it T}$_{c}$ varies between 12 and 24 K,
depending on synthesis conditions, and may be raised to 40 K by long-term
annealing. The temperature dependence of the normal-state resistivity is
again characteristic of a strongly underdoped superconducting cuprate
compound. The ferromagnetic transition at {\it T}$_{Curie}$=133 K causes
only a small yet noticeable drop in the resistivity indicating that the RuO$%
_{2}$ layer is almost insulating above {\it T}$_{Curie}$ while being poorly
conducting in the ferromagnetic state \cite{5}.
\subsection{The technique of muon spin rotation}
The muon spin rotation ($\mu $SR) experiments have been performed at the M15
beamline of TRIUMF in Vancouver, Canada, which provides 100\% spin polarized
muons. The $\mu $SR technique is especially suited for the study of magnetic
materials and allows one to study the homogeneity of the magnetic state on a
microscopic scale and also to access its volume fraction \cite{11}. The $\mu 
$SR technique typically covers a time window of 10$^{-6}$ to 10$^{-9}$
seconds and allows one to detect internal magnetic fields over a wide range
of 0.1G to several Tesla. The 100\% spin-polarised `surface muons' (E$_{\mu
} $ $\approx $ 4.2 MeV) are implanted into the bulk of the sample where they
thermalize very rapidly ($\sim $10$^{-12}$s) without any noticeable loss in
their initial spin polarization. Each muon stops at a well-defined
interstitial lattice site and, for the perovskite compounds, forms a muoxyl
bond with one of the oxygen atoms \cite{13}. The whole ensemble of muons is
randomly distributed throughout a layer of 100-200 $\mu $m thickness and
therefore probes a representative part of the sample volume. Each muon spin
precesses in its local magnetic field B$_{\mu }$ with a precession frequency
of, $\nu _{\mu }$ = ($\gamma _{\mu }/2\pi )$ $\cdot $\ B$_{\mu }$, where $%
\gamma _{\mu }$/2$\pi $= 135.5 MHz/T is the gyromagnetic ratio of the
positive muon. The muon decays with a mean life time of $\tau _{\mu ^{+}}$ $%
\approx 2.2$\ $\mu $s$^{-1}$ into two neutrinos and a positron which is
preferentially emitted along the direction of the muon spin at the instant
of decay. The time evolution of the spin polarization {\it P}({\it t}) of
the muon ensemble can therefore be obtained via the time-resolved detection
of the spatial asymmetry of the decay positron emission rate. More details
regarding the zero-field (ZF) $\mu $SR technique are given below.

\section{Experimental results}
\subsection{DC magnetization}
Before we discuss the result of the $\mu $SR experiments, we first present
some DC-magnetization data which establish that the RuSr$_{2}$GdCu$_{2}$O$%
_{8}$ sample exhibits a spontaneous magnetization at a ferromagnetic
transition of {\it T}$_{Curie}$ = 133 K and becomes superconducting at a
much lower temperature {\it T}$_{c}$ = 16 K. Figure 3(a) shows the
temperature dependence of the volume susceptibility, $\chi _{V}$, which has
been obtained after zero-field cooling the sample to T = 2 K, then applying
an external field of H$^{ext}$ = 5.5 Oe, and subsequently warming up to T =
200 K. The density of the sample has been assumed to be $\rho $\ = 6.7 g/cm$%
^{3}$ corresponding to stoichiometric RuSr$_{2}$GdCu$_{2}$O$_{8}$ with
lattice parameters of a = 3.84 \AA\ and c = 11.57 \AA\ \cite{8}. 
The superconducting transition is evident in Fig 3(a) from the onset of a
pronounced diamagnetic shift below {\it T}$_{c}$ = 16 K. The diamagnetic
shift at the lowest available temperature of {\it T} = 2 K corresponds to an
almost complete diamagnetic shielding of the sample volume, implying that at
least the surface region of the sample is homogeneously superconducting. In
fact, all pieces that have been cut from the pellet exhibit a similarly
large diamagnetic shielding effect (small differences can be attributed to
different demagnetization factors).
Nevertheless, the DC-magnetisation
measurements cannot give unambiguous evidence for the presence of bulk
superconductivity since an almost complete diamagnetic shielding may also be
caused by a filamentary structure of superconducting material in a small
fraction of the otherwise non-superconducting material. 

\begin{figure}
\epsfxsize=5.3 cm
\epsfbox{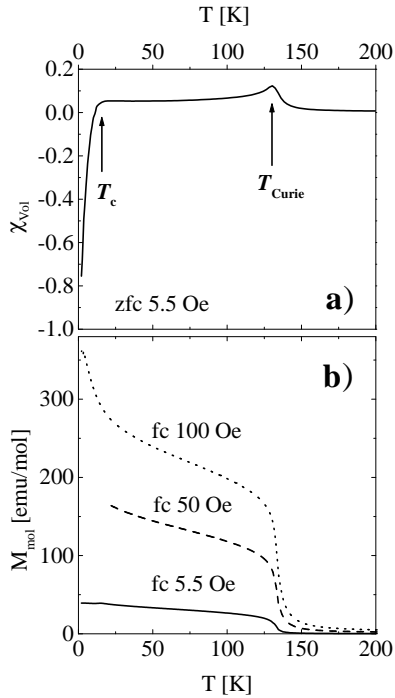}
\caption{(a) The temperature-dependence of the zero-field cooled DC volume
magnetization, $\chi _{V}$, of RuSr$_{2}$GdCu$_{2}$O$_{8}$. The arrows show
the superconducting- and the ferromagnetic transition at {\it T}$_{c}$=16 K
and {\it T}$_{Curie}$ = 133 K, respectively. (b) The field-cooled molar
magnetization, {\it M}$_{mol}$, for applied fields of {\it H}=5.5, 10, 100
Oe.}
\end{figure}

We note however,
that unequivocal evidence for the occurrence of bulk superconductivity in
RuSr$_{2}$GdCu$_{2}$O$_{8}$ has recently been obtained from specific heat
measurements which reveal a sizeable jump of $\Delta \gamma \equiv
Cp/T\approx 0.35$ mJ/g at.K$^{2}$ at T$_{c}$, comparable to or greater than
that seen in other underdoped cuprates \cite{10}. For comparison in strongly
underdoped YBa$_{2}$Cu$_{3}$O$_{7-\delta }$ it is found that $\Delta \gamma
\approx 0.2-0.3$ mJ/g at.K$^{2}$ \cite{14}. We also note that the specific
heat measurements have been performed on the same samples which have been
studied by $\mu $SR- and DC-magnetisation measurements. Figure 3(b) displays
the (low) field-cooled molar magnetization {\it M}$_{m}$ for applied fields
of {\it H}$^{ext}$ = 5.5, 50 and 100 Oe. The ferromagnetic transition at 
{\it T}$_{Curie}$ = 133 K is evident from the sudden onset of a spontaneous
magnetization. Evidently, the magnetic order parameter has at least a
sizeable ferromagnetic component and it persists almost unchanged to the
lowest measured temperature T = 2 K. In particular, it does not appear to
weaken as superconductivity sets in at {\it T}$_{c}$ = 16 K. Additional
evidence for the presence of ferromagnetic order is presented in Fig. 4,
which shows that the isothermal magnetization loops at {\it T} = 5 K and 50
K exhibit hysteretic magnetic behavior with a remanent magnetization {\it M}$%
_{rem}$ $\approx $ 400 Oe at 5 K and 200 Oe at 50 K.

\begin{figure}
\epsfxsize=6.6 cm
\epsfbox{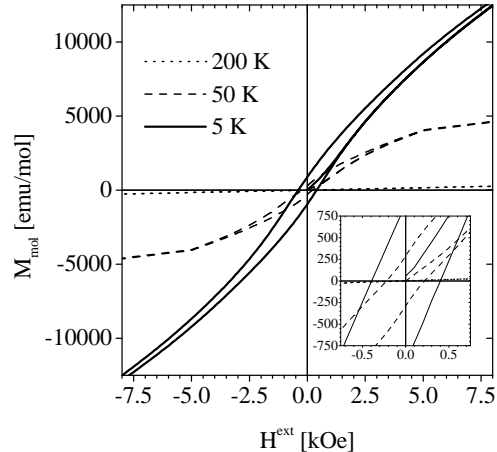}
\caption{The isothermal magnetization loops of RuSr$_{2}$GdCu$_{2}$O$_{8}$ at%
{\it \ T} = 5, 50 and 200 K. The inset shows a magnification of the
low-field region.}
\end{figure}

Having established the existence of ferromagnetic order, the question arises
of whether it involves the Ru-moments or the Gd-moments. In the following we
present high-temperature susceptibility data which indicate that the
ferromagnetic order involves only the Ru$-$moments, whereas the Gd-moments
remain in the paramagnetic state below {\it T}$_{Curie}$. Figure 5 shows the
inverse molar susceptibility, 1/$\chi _{m}$ $\approx $ ({\it M}$_{m}$/{\it H}%
$^{ext}$)$^{-1}$\ obtained for different external fields in the range 5.5 $%
\leqslant ${\it \ H}$^{ext}$ $\leqslant $ 1000 Oe (solid lines) in the
temperature region 200 K%
%TCIMACRO{\TEXTsymbol{<}}%
%BeginExpansion
\mbox{$<$}%
%EndExpansion
{\it T}%
%TCIMACRO{\TEXTsymbol{<}}%
%BeginExpansion
\mbox{$<$}%
%EndExpansion
400 K. Shown by the plus signs (+) is the best fit to the experimental data
using a two-component `Curie-Weiss + Curie-function', $\chi $ = C$_{1}$/(%
{\it T}-$\Theta $) + C$_{2}$/{\it T} , with $\Theta $\ = {\it T}$_{Curie}$ =
133 K kept fixed. This function describes the experimental data rather well
and it gives us very reasonable values for the magnetic moments, with $\mu
_{1}$ = 1.05(5) $\mu _{B}$ for the moments that order at {\it T}$_{Curie}$
and \ $\mu _{2}$ = 7.4(1) $\mu _{B}$ for the moments that remain
paramagnetic below {\it T}$_{Curie}$. The magnetic moment of the
paramagnetic component agrees reasonably well with the expected magnetic
moment of Gd$^{3+}$ which for a free Gd$^{3+}$-ion \cite{15} is $\mu $(Gd$%
^{3+}$) = 7.94 $\mu _{B}$ and $\mu $(Gd$^{3+}$) = 7.4 $\mu _{B}$ for the
structurally-similar GdBa$_{2}$Cu$_{3}$O$_{7-\delta }$\ compound \cite{16}.
On the other hand, the value of the Ru-moments with $\mu $(Ru) = 1.05(5) $%
\mu _{B}$ also appears to be reasonable. For Ru$^{5+}$\ the number of 4d
electrons is 3 and the free ion value of the magnetic moment is 3 $\mu _{B}$
for the high spin state and 1 $\mu _{B}$ for the low spin state. The
experimentally observed value of $\mu $(Ru) = 1.05(5) $\mu _{B}$ therefore
seems to imply that Ru$^{5+}$ is in the low spin state. Shown in the inset
of Fig. 5 is the field-dependent magnetization for the temperatures {\it T}
= 2, 30, 50, 100 and 300 K. The low-temperature magnetization can be seen to
saturate at a value of $\mu _{sat}$ $\approx $\ 8 $\mu _{B},$ as may be
expected for a system that contains one Gd-moment per formula unit with $\mu 
$(Gd) = 7 $\mu _{B}$ plus one Ru-moment with $\mu $(Ru) = 1 $\mu _{B}$.
\begin{figure}
\epsfxsize=5.6 cm
\epsfbox{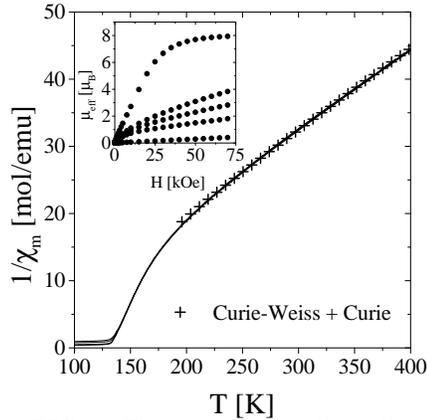}
\caption{The temperature-dependent inverse molar susceptibility, 1/$\chi _{m}$
for the high temperature range of 400 K \mbox{$>$}% {\it T} \mbox{$>$}%
100 K. The plus signs show the best fit using a two component `Curie-Weiss-
+ Curie-function'. Shown in the inset is the saturation magnetization in
units of effective Bohr magnetons per unit volume as a function of applied
field at temperatures of {\it T} = 2, 30, 50, 100, 200 K.}
\end{figure}
\begin{figure}
\epsfxsize=5.5 cm
\epsfbox{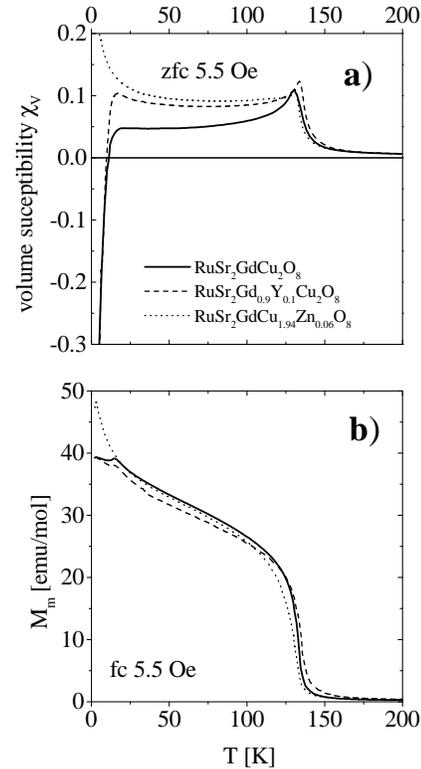}
\caption{(a) The temperature-dependent volume susceptibility, $\chi _{V}$, of
RuSr$_{2}$GdCu$_{2}$O$_{8}$ (solid line), for Zn-substituted RuSr$_{2}$GdCu$%
_{1.94}$Zn$_{0.06}$O$_{8}$ (dotted line) and for Y $\longleftrightarrow $ Gd
cosubstituted RuSr$_{2}$Gd$_{0.9}$Y$_{0.1}$Cu$_{2}$O$_{8}$ (dashed line).
(b) The temperature-dependent molar magnetization {\it M}$_{m}$, shown by
the same symbols as in (a).}
\end{figure}
The idea that the Gd-moments do not participate in the ferromagnetic
transition at {\it T}$_{Curie}$\ = 133 K is supported by the result of
DC-magnetization measurements on the 10\% Y $\leftrightarrow $\ Gd
cosubstituted RuSr$_{2}$Gd$_{0.9}$Y$_{0.1}$Cu$_{2}$O$_{8}$. Figures 6(a) and
6(b) display the zero-field-cooled- and the field-cooled susceptibilities
(dashed lines) and compare them with the corresponding data on the pure RuSr$%
_{2}$GdCu$_{2}$O$_{8}$ sample (solid line). It is evident that the
ferromagnetic transition is not significantly affected by the partial
substitution of non-magnetic Y$^{3+}$ for magnetic Gd$^{3+}$. Also shown in
Fig. 6(a) and 6(b) by the dotted lines are the results for the
Zn-substituted RuSr$_{2}$Cu$_{1.94}$Zn$_{0.06}$O$_{8}$ sample{\bf .} The
circumstance that the ferromagnetic order is not affected by the
Zn-substitution supports our view that the majority of the Zn-impurities has
been introduced into the CuO$_{2}$ layers while hardly any of them reside
within the RuO$_{2}$-layers. Moreover, we infer from the rapid {\it T}$_{c}$%
-suppression upon Zn-substitution that only the CuO$_{2}$ layers host the
superconducting charge carrier in RuSr$_{2}$GdCu$_{2}$O$_{8}$.

\subsection{Zero-field muon-spin-rotation (ZF-$\protect\mu $SR)}
Next we discuss the result of the zero-field (ZF) $\mu $SR experiments.
Figures 7(a) and 7(b) show representative ZF-$\mu $SR spectra for the
evolution of the normalized time-resolved muon spin polarization {\it P}(%
{\it t})/{\it P}(0) at temperatures of {\it T} = 5 and 48 K. 

\begin{figure}
\epsfxsize=5.8 cm
\epsfbox{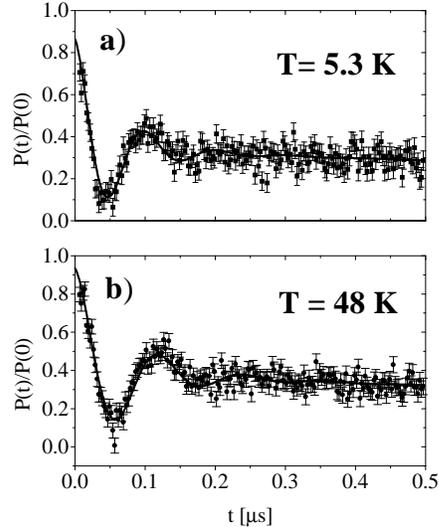}
\caption{The time-resolved normalized muon-spin polarization, {\it P}({\it t}%
)/{\it P}({\it t}=0), at temperatures of (a) {\it T} = 5.3 K 
\mbox{$<$}%
{\it T}$_{c}$ and (b) {\it T}$_{c}$ 
\mbox{$<$}%
{\it T} = 48 K 
\mbox{$<$}%
{\it T}$_{Curie}$ = 133 K. The large oscillatory component gives clear
evidence for the presence of a homogeneous magnetically-ordered state.}
\end{figure}

The value of
the initial muon spin polarization, P(0), has been determined by a
transverse field (TF) $\mu $SR experiment performed on the same sample at a
temperature above{\it \ T}$_{Curie}$. In the ferromagnetic state below {\it T%
}$_{Curie}$=133 K we find that the spectra are well described by the
relaxation function:

$\qquad \qquad {\it P}({\it t})/{\it P}({\it t=}0)=A_{1}\cdot exp(-\lambda 
{\it t})\cdot cos(2\pi \langle \nu _{\mu }\rangle {\it t})+A_{2}\cdot
exp(-\Lambda {\it t}),\qquad (1)$\newline
where $\langle \nu _{\mu }\rangle $ is the average muon spin precession
frequency which corresponds to the average value of the spontaneous internal
magnetic field at the muons sites, $\langle \nu _{\mu }\rangle $ = $\gamma
_{\mu }/2\pi $ $\cdot $\ $\langle ${\it B}$_{\mu }\rangle $, with $\gamma
_{\mu }$\ = 835.4 MHz/T the gyromagnetic ratio. The damping rate of the
non-oscillating (longitudinal) component, $\Lambda ,$ is proportional to the
dynamic spin-lattice relaxation rate, $\Lambda $ $\sim $ 1/T$_{1}$, whereas
the relaxation rate of the oscillating (transverse)\ component, $\lambda $,
is dominated by the static distribution of the local magnetic field, i.e., $%
\lambda \approx \gamma _{\mu }\cdot \langle \Delta {\it B}_{\mu }\rangle $.
Figure 8 shows the temperature dependence of (a) the precession frequency, $%
\langle \nu _{\mu }\rangle ({\it T}),$ (b) the transverse relaxation rate, $%
\lambda ({\it T})$, and (c) the longitudinal relaxation rate, $\Lambda ({\it %
T})$.

\begin{figure}
\epsfxsize=6 cm
\epsfbox{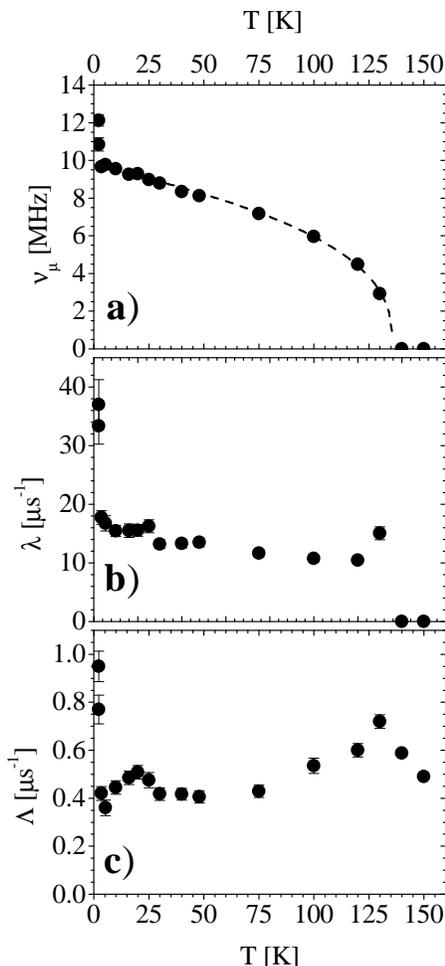}
\caption{The temperature dependence of the $\mu $SR signal of RuSr$_{2}$GdCu$%
_{2}$O$_{8}$ (a) The muon spin precession frequency, $\nu _{\mu }$({\it T})
[MHz] = 135.5 [MHz/T] $\langle {\it B}_{\mu }\rangle .$\ Shown by the dashed
line is the best fit using the scaling function $\nu _{\mu }(T)=\nu
_{o}(1-T/T_{Curie})^{\beta }$. with $\beta $ = 0.333(5) {\it T}$_{Curie}$ =
133(1) K, and $\nu _{o}=9.70(5)$ MHz ({\it B}$_{\mu }$ = 720(10) G). (b) The
relaxation rate of the precessing component, $\lambda $({\it T}). (c) The
relaxation rate of the non-precessing component, $\Lambda $({\it T}) $\sim $
1/T$_{1}$.}
\end{figure}

Before we discuss the ZF-$\mu $SR data in more detail, we first emphasize
the most important implications, which are evident from Figs. 7 and 8.
Firstly, the presence of an oscillating component in the ZF-$\mu $SR spectra
for {\it T}\ 
%TCIMACRO{\TEXTsymbol{<} }%
%BeginExpansion
\mbox{$<$}%
%EndExpansion
{\it T}$_{Curie}$ = 133 K gives unambiguous evidence for an ordered magnetic
state which is homogeneous on a microscopic length scale (of typically 20
\AA ). Secondly, from the amplitude of the oscillating component (A$%
_{1}\approx $2/3) we can deduce that the magnetically-ordered state accounts
for more or less the entire volume of the sample. And thirdly, from the
temperature dependence of the $\mu $SR signal it becomes clear that the
magnetic order persists almost unchanged in the superconducting state.

\subsubsection{The volume fraction of the magnetic phase}
In the following we outline how the volume fraction of the
magnetically-ordered phase is obtained from the amplitude of the oscillating
component of the ZF-$\mu $SR spectra. For a polycrystalline sample with
randomly-orientated grains in zero external field the local magnetic field,
on average, is parallel (perpendicular) to the direction of the muon spin
direction with probability 1/3 (2/3). For a homogeneous magnetically-ordered
sample one therefore expects that 2/3 of the amplitude of the ZF-$\mu $SR
signal (the transverse component) exhibit an oscillatory behavior, while 1/3
of the signal (the longitudinal component) is non-oscillating and only
slowly damped due to spin-flip excitations. On the other hand, for a sample
with inhomogeneous magnetic order, for example containing non-magnetic
regions, the amplitude of the oscillating signal will be accordingly reduced
and a second non-oscillating transverse component will appear. If the
non-magnetic regions are microscopically small, this non-oscillating
component is likely to have a rather large damping rate of the order of $%
\lambda $ $\sim $ $\gamma _{\mu }\cdot \langle B_{\mu }\rangle $ due to
stray fields which are imposed by the neighboring magnetic domains. From
Fig. 7(a) and 7(b) it can be seen that the ZF-$\mu $SR data on RuSr$_{2}$GdCu%
$_{2}$O$_{8}$ give no indication for such an inhomogeneous magnetic state.
As was mentioned above, the amplitude of the initial muon spin polarization 
{\it P}({\it t} = 0) has been determined from a transverse-field (TF)-$\mu $%
SR measurement. From the size of the amplitude of the oscillatory component
we deduce that more than 80 \% of the sample is magnetically ordered below 
{\it T}$_{Curie}$ = 133 K. Based on this analysis we estimate that the
volume fraction of any disordered magnetic- or non-magnetic phase must be
well below 20\%. Note that some of the muons (typically 10-20 \%) do not
stop inside the sample but somewhere in the cryostat walls. In the ZF-$\mu $%
SR experiment these muons give rise to a missing fraction since their
spin-polarization is much more slowly damped than for the rest of the
signal. In the TF-$\mu $SR experiment, however, this very slowly damped
component can be detected via its precession in the external field and it
contributes to the total muon spin polarization P(0).\ The 80\% fraction of
the magnetically ordered phase therefore has to be regarded as a lower
bound. In fact, it is rather likely that the entire sample volume is
magnetically ordered. Finally, we note that the muons apparently occupy only
one muon site, since only one precession frequency is seen in the ZF-$\mu $%
SR spectra. Also it is clear from the ZF-$\mu $SR data that muon diffusion
effects are negligibly small below {\it T}$_{Curie}$ = 133 K, similar to the
other cuprate superconductors where muon diffusion is observed only at
significantly higher temperatures of T $\geq $ 250 K \cite{13}.

\subsubsection{Local magnetic field at the muon site}
It is evident from Fig. 8(a) that the muon spin precession frequency (the
local field at the muon site) does not exhibit any strong anomaly at the
superconducting transition temperature {\it T}$_{c}$. Instead, as shown by
the dashed line, the temperature of the muon spin precession frequency $%
\langle \nu _{\mu }\rangle ({\it T})$\ (and thus of the magnetic order
parameter) is well described by the function $\nu _{\mu }({\it T})$ = $\nu
_{o}\left( 1-{\it T}/{\it T}_{c}\right) ^{\beta },$\ with $\nu _{o\bigskip }$
= 9.7(1) MHz (corresponding to $\langle B_{\mu }\rangle (T\rightarrow
0)\approx $\ 720(10) G), {\it T}$_{c}$ = 133(1) K, and $\beta $\ =
0.333(5).\ This functional form is strictly valid only in the critical
regime close to {\it T}$_{Curie}$ but it can be seen to provide a reasonable
description of the magnetic order parameter over a fairly wide temperature
range of {\it T}$_{Curie}$ $\geqslant $\ {\it T} $\geqslant $ 5 K. The
anomaly at very low temperature arises most likely from the magnetic
ordering transition of the Gd-moments at {\it T}$_{N}\approx $2.6 K. Note
that for the structurally related compound GdBa$_{2}$Cu$_{3}$O$_{7-\delta }$
(Gd-123) the antiferromagnetic ordering transition of the Gd-moment occurs
at a very similar temperature of {\it T}$_{N}$=2.3K [16,17]. The value of
the critical exponent $\beta $ = 0.333 is close to the theoretical value
0.345 in the 3D XY model \cite{18}. We cannot determine with certainty the
number of components in the spin system with these data and, in particular,
distinguish between the 2-component XY ($\beta $= 0.345) and the 3-component
Heisenberg ($\beta $ = 0.365) models. The contribution of ferromagnetic
fluctuations above {\it T}$_{Curie}$ to the susceptibility provides better
discrimination as will be discussed later.

The oscillating transverse component exhibits a damping rate of the order of 
$\lambda $ $\approx $\ 10-15 $\mu $s$^{-1}$ corresponding to a spread in the
local magnetic field of $\langle \Delta {\it B}_{\mu }\rangle /\langle {\it B%
}_{\mu }\rangle \approx $ 0.2. This 20 \% spread of the local magnetic field
does not seem to agree with a scenario where the ferromagnetic order is
assumed to exhibit a spiral modulation (with a wavelength shorter than the
superconducting coherence length of typically 20 \AA\ in the cuprates)
and/or to be spatially inhomogeneous as in ErRh$_{4}$B$_{4}$ \cite{19}, HoMo$%
_{6}$S$_{8}$ \cite{20} and Y$_{9}$Co$_{7}$ \cite{21}. Instead, we emphasize
that the observed spread in the local magnetic field can be accounted for by
the grain boundary effects and by the differences in the demagnetization
factors of the individual grains which naturally arise for a polycrystalline
sample that has a very small average grain size of about 1 $\mu $m \cite{8}.
Also, we point out that recent transmission-electron-microscopy (TEM)
studies have revealed that our present Ru-1212 sample contains [100]
rotation twins and also exhibits some cationic disorder due to the
intermixing of Sr $\leftrightarrow $\ Gd and to a lesser extent of Ru $%
\leftrightarrow $\ Cu \cite{8}. These kinds of structural imperfections
certainly tend to further increase the transverse relaxation rate $\lambda $
of the ZF-$\mu $SR spectra. Meanwhile, we have prepared Ru-1212 samples
which are structurally more perfect (by sintering at slightly higher
temperature and for longer periods) \cite{8}. Recent DC-magnetization
measurements have shown that these crystallographic defects do not affect
the fundamental magnetic and superconducting behavior. In fact, both the
superconducting- and the ferromagnetic transitions become somewhat sharper
and T$_{c}$ and T$_{Curie}$ are slightly increased for these structurally
more perfect samples \cite{8}. Additional $\mu $SR measurements on these
samples are presently under way.

\subsubsection{\protect\bigskip Longitudinal relaxation rate, $\Lambda $ $%
\sim $ 1/T$_{1}$ }
The temperature dependence of the relaxation rate of the non-oscillating
component of the ZF-$\mu $SR signal, ${\it \Lambda }$({\it T}) $\sim 1/T$ ,
is shown in Fig. 8(c). As a function of decreasing temperature ${\it \Lambda 
}({\it T})$\ can be seen to exhibit a cusp-like feature at the ferromagnetic
transition of the Ru-moments at {\it T}$_{Curie}$=133 K and a step-like
increase at very low temperature which most likely is related to the
ordering of the Gd moments. The cusp feature at {\it T}$_{Curie}$ = 133 K
characterizes the slowing down of the spin dynamics of the Ru-moments as the
ferromagnetic transition is approached. The cusp maximum occurs when the
spin fluctuation rate, $\tau _{c}$, equals the typical $\mu $SR time scale
for $\tau _{c}$ $\sim $\ 10$^{-6}$ \cite{11}. Note, that in the
ferromagnetically-ordered state the longitudinal relaxation rate remains
unusually large with values of ${\it \Lambda }$({\it T} 
%TCIMACRO{\TEXTsymbol{<}}%
%BeginExpansion
\mbox{$<$}%
%EndExpansion
%TCIMACRO{\TEXTsymbol{<} }%
%BeginExpansion
\mbox{$<$}%
%EndExpansion
{\it T}$_{Curie})$ $\approx $\ 0.3-0.4 $\mu $s$^{-1}$ that are at least an
order of magnitude larger than expected for a classical ferromagnet \cite{22}%
\ (where two-magnon excitations provide the major contribution to spin
dynamics). We have confirmed by a $\mu $SR measurements in a longitudinal
field of H$^{LF}$=6 kOe that this large relaxation rate is indeed
characteristic for the longitudinal component of the $\mu $SR signal. At
present we cannot provide a definite explanation of the origin of the
unusually large value of $\Lambda $. However, we emphasize that the RuSr$%
_{2} $GdCu$_{2}$O$_{8}$ system can be expected to exhibit a rather complex
magnetic behavior since, besides the ferromagnetically-ordered Ru moments,
it also contains the larger Gd moments with $\mu (Gd^{3+})$ $\approx $ 7.4 $%
\mu _{B}$ which remain paramagnetic below {\it T}$_{Curie}$. The magnetic
ordering transition of the Gd moments at T$\approx $2.6 K is evident in the
ZF-$\mu $SR data in Fig. 8(a)-(c) from the sudden increase in the local
magnetic field (or the $\mu $SR precession frequency, $\langle \nu _{\mu
}\rangle )$ and a corresponding increase in both relaxation rates, ${\it %
\lambda }$\ and ${\it \Lambda }$. In addition, we note that recently it has
been shown by $\mu $SR measurements that in strongly underdoped high-T$_{c}$
cuprate superconductors (like the present RuSr$_{2}$GdCu$_{2}$O$_{8}$
compound) also the Cu moments exhibit a spin-glass type freezing transition
at low temperature \cite{23}. Finally, it appears that the longitudinal
relaxation rate $\Lambda $ exhibits an additional weak anomaly at a
temperature of {\it T}$\approx $20 K, i.e. in the vicinity of the
superconducting transition at T$_{c}$=16 K. At present we are not sure
whether this effect is related to the onset of superconductivity. From Fig.
8(b) it appears that the transverse relaxation rate also exhibits a steplike
increase in the same temperature range. The local magnetic field at the muon
site, however, (see Fig. 8a) does not seem to exhibit any anomaly in the
vicinity of T$_{c}$. We expect that further $\mu $SR measurements on rare
earth (RE) substituted RuSr$_{2}$Gd$_{1-x}$RE$_{x}$Cu$_{2}$O$_{8}$ samples,
as well as on less strongly underdoped samples with higher critical
temperatures of T$_{c}$ up to 40 K, should shed more light on the complex
magnetic behavior and its interplay with superconductivity in the Ru-1212
system.

\subsection{Dipolar Field Calculation}
While the ZF-$\mu $SR data give clear evidence for the presence of a
homogeneous magnetically ordered state, they do not provide any direct
information about the origin of the magnetic moments, the type of the
magnetic order and its direction. Based on dipolar-field calculations of the
local magnetic field at the muon site, however, one can test the consistency
with an assumed magnetic structure. The result of these calculations depends
on the location of the interstitial muon site and also on the orientation of
the Ru-moments. Unfortunately, for the Ru-1212 system neither of these is
accurately known at present. Nevertheless, it seems plausible that the muon
site is similar to that in YBa$_{2}$Cu$_{3}$O$_{7-\delta }$ (and other
related cuprate compounds) where the positive muon forms a hydroxyl bond
with the apex oxygen and is located at the so-called `apical-site' near the
point (0.12a,0.225b,0.14c)\cite{13}. Indeed, as is summarized in table 1, we
obtain rather good agreement with the experimental value of $\langle {\it B}%
_{\mu }\rangle $({\it T}$\rightarrow $0) = 720 G if we take a similar
apical-site near the point (0.13a, 0.22b, 0.16-0.17c) and assume that the
ferromagnetically-ordered Ru-moments ($\mu $(Ru)=1 $\mu _{B})$ are oriented
along the RuO$_{2}$ plane either along the Ru-O bond, [100], or along the
diagonal [110] (see table 1). For the [110] orientation, however, there
exist two magnetically inequivalent muon sites which should give rise to two
distinctive precession frequencies in the $\mu $SR spectra (which are not
observed experimentally). For the Ru-moments oriented perpendicular to the
RuO$_{2}$ layer along [001] the resulting local magnetic field at the
apex-site is significantly larger than the experimental value. In order to
obtain reasonable agreement with experiment for the [001]-orientation one
has to assume that the muon site is located much closer to the CuO$_{2}$
planes. Such a muon site, however, is not very realistic (simply speaking
the positive muon is repelled by the positively charged CuO$_{2}$ planes)
and has not been observed in any of the related cuprate compounds. We thus
tentatively conclude that the moments align in-plane consistent with the
2-component XY scenario. While this result is rather convenient in terms of
the coexistence of the ferromagnetic order of the Ru moments and the
superconductivity which resides within the CuO$_{2}$ layers as discussed
below, one has to keep in mind that the underlying assumptions are rather
crude. For more detailed and decisive information on the structure and the
orientation of the Ru spin order we must await the result of neutron
scattering experiments.

\section{A possible scenario for coexistence of ferromagnetic- and
superconducting order}
Having established that the ferromagnetic and the superconducting order
parameter coexist on a microscopic scale, we arrive at the important
question as to how this system manages to avoid strong pair-breaking
effects. We suspect that the answer is closely related to the layered
structure of the hybrid ruthenate-cuprate compound and, in particular, to
the purely two-dimensional coherent charge transport in the strongly
underdoped CuO$_{2}$ planes. We envisage a scenario where the
ferromagnetically ordered Ru spins are aligned in the RuO$_{2}$ plane having
a very large out-of-plane anisotropy while the charge dynamics of the
superconducting CuO$_{2}$ planes is purely two-dimensional, i.e., coherent
charge transport occurs only along the direction of the CuO$_{2}$ planes.
For such a configuration the principal pair-breaking effect due to the
electromagnetic interaction can be minimized, since the dot product of the
magnetic vector potential (which then is normal to the planes) and the
momentum of the Cooper pair (which is parallel to the planes) vanishes. An
additional requirement is that the direct hyperfine interaction between the
superconducting electrons of the CuO$_{2}$ planes and the ordered Ru spins
has to be extremely small. Both requirements may be fulfilled in the present
Ru-1212 system due to the confinement of the superconducting electrons of
the strongly underdoped CuO$_{2}$ planes. The absence of magnetic
pairbreaking is suggested by the fact that the {\it T}$_{c}$ value is fully
consistent with the underdoped state indicated by the thermoelectric power 
\cite{12}. So far we have been unable to significantly increase the doping
state (by e.g. Ca substitution) so as to explore these implications.
Furthermore, we have not yet succeeded in crystallographically aligning
powders or growing single crystals which would allow one to investigate the
magnetic anisotropy. However, further information can be obtained by
examining the ferromagnetic fluctuations above {\it T}$_{Curie}$ as seen in
the divergence of the susceptibility. Figure 9 shows d$\chi $/d{\it T}
plotted versus ({\it T}/{\it T}$_{Curie}$-1) for the zero-field-cooled
susceptibility for {\it T} 
%TCIMACRO{\TEXTsymbol{>} }%
%BeginExpansion
\mbox{$>$}%
%EndExpansion
{\it T}$_{Curie}$. The slope of -2.30(3) indicates a critical exponent of $%
\gamma $\ = 1.30(3) consistent with 3D XY fluctuations for which $\gamma $\
= 1.32 \cite{18} and again consistent with orientation of the Ru-moments
within the a-b plane.

\begin{figure}
\epsfxsize=6 cm
\epsfbox{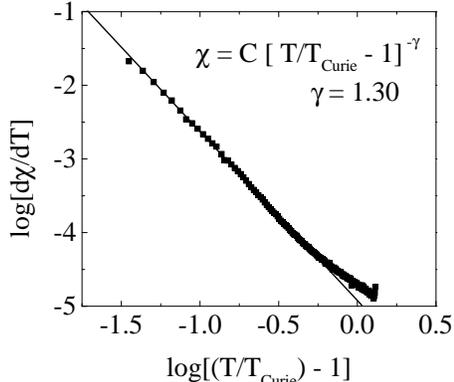}
\caption{d$\chi $/d{\it T} plotted versus ({\it T}/{\it T}$_{Curie}$-1) for
the zero-field cooled susceptibility in the temperature range of {\it T} 
\mbox{$>$}%
{\it T}$_{Curie}$. The slope of -2.30(3) indicates a critical exponent of $%
\gamma $\ = 1.30(3) consistent with 3D XY fluctuations for which $\gamma $\
= 1.32 \cite{18} and consistent with orientation of the Ru-moments within
the a-b plane.}
\end{figure}

Finally, we note that an alternative (and highly speculative) explanation
for the coexistence of high-T$_{c}$ superconductivity and ferromagnetic
order in the present Ru-1212 superconductor could be that the
superconducting order parameter has a non-zero angular momentum which itself
breaks time-reversal symmetry. Such a highly unconventional order parameter
symmetry has been discussed also in the context of the Sr$_{2}$RuO$_{4}$
superconductor. We\ point out, however, that at present we have no evidence
in favor of such a scenario.

\section{Conclusions}
In summary, we have performed DC-magnetization and zero-field muon spin
rotation (ZF-$\mu $SR) measurements which characterize the superconducting-
and the magnetic properties of the hybrid cuprate-ruthenate compound RuSr$%
_{2}$GdCu$_{2}$O$_{8}$. The DC-magnetization data establish that this
material exhibits ferromagnetic order (or at least magnetic order with a
sizeable ferromagnetic component) below {\it T}$_{Curie}$ = 133 K and
becomes superconducting at a much lower temperature of {\it T}$_{c}$ = 16 K.
We obtain evidence that superconducting charge carriers originate from the
CuO$_{2}$ planes, while the ferromagnetic order is associated with the Ru
moments with $\mu $(Ru) $\approx $ 1 $\mu _{B}$. The larger Gd moments with $%
\mu $(Gd) $\approx $\ 7.4 $\mu _{B}$ do not appear to participate in the
ferromagnetic transition but remain paramagnetic to very low temperature and
undergo most likely an antiferromagnetic transition at {\it T}$_{N}$=2.6 K.
The ZF-$\mu $SR experiments provide evidence that the ferromagnetic phase is
homogeneous on a microscopic scale and accounts for most of the sample
volume. Furthermore, they indicate that the magnetically ordered state is
not significantly modified by the onset of superconductivity. This rather
surprising result raises the question as to how ferromagnetic and
superconducting order can coexist on a microscopic scale while avoiding
strong pairbreaking effects that tend to destroy superconductivity. We have
outlined a possible scenario which relies on the two-dimensional charge
dynamics of the CuO$_{2}$ planes and the assumption that the ferromagnetic
order parameter of the Ru-moments is confined to the RuO$_{2}$ layers.

\section{ Acknowledgments}
We would like to acknowledge S. Kreitzman and B. Hitti for the technical
support during the $\mu $SR experiments at TRIUMF. JLT thanks the Royal
Society of New Zealand for financial support under a James Cook Fellowship.
ChN and ThB thank the German BMBF for financial support. CES and DRN were
supported by US Air Force Office of Scientific Research grant
F49620-97-1-0297.
\end{multicols}
* Permanent address: Institute of Experimental Physics, Warsaw University,
Ho\.{z}a 69, 00-681 Warsaw, Poland;
 
\# Present address: 1318 Tenth St., Saskatoon, Canada S7H OJ3

Table 1: The local magnetic field at the muon site, $\langle B_{\mu }\rangle
,$ obtained from the dipolar-field calculation. The results are shown for
two different muon sites and for three different orientations with the Ru
moments ($\mu $(Ru$^{5+}$)=1 $\mu _{B})$ ferromagnetically ordered along the
Ru-O bond [100], along the diagonal [110], or perpendicular to the RuO$_{2}$
planes [001]. Note that for the [110] orientation there exist two
magnetically inequivalent muon sites.

.

\begin{tabular}{||l||l||l||l||}
\hline\hline
muon-site & [100] & [110] & [001] \\ \hline\hline
(0.13, 0.225, 0.16) & 960 G & 1068/805 G & 1471 G \\ \hline\hline
(0.128, 0.222, 0.175) & 740 G & 824/665 G & 1231 G \\ \hline\hline
\end{tabular}

\end{document}